# A decomposition of Fisher's information to inform sample size for developing fair and precise clinical prediction models - Part 2: time-to-event outcomes

Richard D Riley[1, 2*], Gary S Collins[3], Lucinda Archer[1, 2], Rebecca Whittle[1, 2], Amardeep Legha,[1, 2] Laura Kirton,[4] Paula Dhiman,[3] Mohsen Sadatsafavi,[5] Nicola J. Adderley,[1, 2] Joseph Alderman,[2,6,7] Glen P. Martin[8], Joie Ensor[1, 2]

* Corresponding author: r.d.riley@bham.ac.uk  @Richard_D_Riley

**Author details:**
[1] Department of Applied Health Sciences, School of Health Sciences, College of Medicine and Health, University of Birmingham, Birmingham, UK
[2] National Institute for Health and Care Research (NIHR) Birmingham Biomedical Research Centre, Birmingham, UK.
[3] Centre for Statistics in Medicine, Nuffield Department of Orthopaedics, Rheumatology and Musculoskeletal Sciences, University of Oxford, Oxford, OX3 7LD, UK
[4] Cancer Research UK Clinical Trials Unit, Institute of Cancer and Genomic Sciences, College of Medical and Dental Sciences, University of Birmingham, Birmingham, UK
[5] Respiratory Evaluation Sciences Program, Faculty of Pharmaceutical Sciences, The University of British Columbia, Vancouver, Canada
[6] Institute of Inflammation and Ageing, College of Medicine and Health, University of Birmingham, Birmingham, UK
[7] Queen Elizabeth Hospital, University Hospitals Birmingham NHS Foundation Trust, Birmingham, UK
[8] Division of Informatics, Imaging and Data Science, Faculty of Biology, Medicine and Health, University of Manchester, Manchester Academic Health Science Centre, Manchester, UK

**Funding**: This paper presents independent research supported by an EPSRC grant for 'Artificial intelligence innovation to accelerate health research' (number: EP/Y018516/1), and by the National Institute for Health and Care Research (NIHR) Birmingham Biomedical Research Centre at the University Hospitals Birmingham NHS Foundation Trust and the University of Birmingham. GSC is supported by Cancer Research UK (programme grant: C49297/A27294). PD is supported by Cancer Research UK (project grant: PRCPJT-Nov21\100021). LK is supported by an NIHR Doctoral Fellowship (NIHR303331) and by a core funding grant awarded to the Cancer Research UK Clinical Trials Unit by Cancer Research UK (CTUQQR-Dec22/100006). GPM and RDR are partially supported by funding from the MRC-NIHR Methodology Research Programme [grant number: MR/T025085/1]. RDR and GSC are National Institute for Health and Care Research (NIHR) Senior Investigators.  The views expressed are those of the author(s) and not necessarily those of the NHS, the NIHR or the Department of Health and Social Care.

**Competing Interests**: RDR receives royalties for sales of his textbooks "Prognosis Research in Healthcare" and "Individual Participant Data Meta-analysis". JEA is the co-organiser of Alan Turing Institute Clinical AI interest group (unpaid). NJA received grant funding from the NIHR outside the submitted work.

**Word count**: 8500

**Contributions:** RDR derived the mathematical solutions, initial software code, made the initial applied examples and wrote the paper. JE produced the pmstabilityss module to generalise the code and apply to the examples. All authors provided critical feedback at multiple stages that let to revision of the rationale, methods, applications, fairness checks, and text or figures in the paper. RDR revised the paper after feedback from authors.

**Data availability:** A previous dataset is available of 686 patients diagnosed with node positive breast cancer from 1984 to 1989 and recruited to the German Breast Cancer Study Group (GBSG). We focused on the subset of 220 participants that were ER positive and received tamoxifen treatment. Data is available here https://hbiostat.org/data/

**Human Ethics and Consent to Participate declarations**: not applicable

## Abstract

**Background**

When developing a clinical prediction model using time-to-event data (i.e., with censoring and different lengths of follow-up), previous research focuses on the sample size needed to minimise overfitting and precisely estimating the overall risk. However, instability of individual-level risk estimates may still be large.

**Methods**

We propose using a decomposition of Fisher's information matrix to help examine and calculate the sample size required for developing a model that aims for precise and fair risk estimates. We propose a six-step process which can be used before data collection or when an existing dataset is available. Steps (1) to (5) require researchers to specify the overall risk in the target population at a key time-point of interest; an assumed pragmatic 'core model' in the form of an exponential regression model; the (anticipated) joint distribution of core predictors included in that model; and the distribution of any censoring. The 'core model' can be specified directly or based on a specified C-index and relative effects of (standardised) predictors. The joint distribution of predictors may be available directly in an existing dataset, in a pilot study, or in a synthetic dataset provided by other researchers.

**Results**

We derive closed-form solutions that decompose the variance of an individual's estimated event rate into Fisher's *unit* information matrix, predictor values and total sample size; this allows researchers to calculate and examine uncertainty distributions around individual risk estimates and misclassification probabilities for specified sample sizes. We provide an illustrative example in breast cancer and emphasise the importance of clinical context, including any risk thresholds for decision making, and examine fairness concerns for pre- and post-menopausal women. Lastly, in two empirical evaluations, we provide reassurance that uncertainty interval widths based on our exponential approach are close to using more flexible parametric models.

**Conclusions**

Our approach allows users to identify the (target) sample size required to develop a prediction model for time-to-event outcomes, via the pmstabilityss module. It aims to facilitate models with improved trust, reliability and fairness in individual-level predictions.

## 1. Background

Studies developing a clinical prediction model use a sample of data, ideally representative of a target population (e.g., women diagnosed with breast cancer), to produce a model (e.g., regression, random survival forest, or deep learning) for estimating an individual's risk of a particular outcome (e.g., 5-year risk of breast cancer recurrence) conditional on their values of multiple predictors (covariates, features). An example is the 'Post D-dimer' model of Ensor et al. to estimate an individual's risk (by 1, 2 or 3 years) of a recurrent venous thromboembolism event following cessation of therapy for a first event.[1]

When developing a prediction model there is a responsibility to implement rigorous standards in study design and analysis.[2-4] An important design aspect, which is often neglected,[5-12] is the sample size requirements to develop and evaluate these models. In previous work we outlined how to calculate the minimum sample size needed for model development for time-to-event outcomes,[13] based on (i) estimating the overall event risk by a particular time-point precisely, and (ii) minimising model overfitting for a regression-based prediction model in terms of overall fit and population-level calibration slope. These criteria are implemented in the Stata and R module *pmsampsize*,[14,15] and they aim to target a reliable prediction model at least at the population level, corresponding to the first two model stability levels defined by Riley and Collins.[16]

As clinical prediction models are used to guide *individual* decision making, targeting reliable predictions at the subgroup and individual-level is important, corresponding to the third and fourth stability levels defined by Riley and Collins.[16] This may require considerably larger sample sizes than the minimum recommended by previous approaches targeting the population level. Here, we derive sample size calculations for time-to-event outcomes and targeting precise individual-level predictions, in situations where models are developed to estimate event risk by one or more time-points of interest, using a dataset where individuals will have different lengths of follow-up (e.g., due to censoring). This extends our work for binary outcomes.[16,17] We derive closed-form solutions based on decomposing prediction variances into Fisher's unit information matrix and total sample size, which can be used (either before data collection or when an existing dataset is available) to examine how sample size impacts uncertainty distributions and intervals around individual-level risk estimates from a model with a pre-specified core set of predictors. Our aim is to help researchers identify a sample size (before any model building) that targets precise individual-level risk estimates for informing patient counselling, enhancing clinical decision making and ensuring fairness.[18]

The article outline is as follows. In Section 2 we summarise our existing sample size approach targeting precise population-level predictions. In Section 3 we outline our new proposal to examine sample size requirements to target precise and fair individual-level predictions, based on analytic solutions assuming an exponential survival model. We also show how this allows the probability of misclassification to be examined in situations where risk thresholds are relevant for decision making. Section 4 applies our proposal to a clinical example. Section 5 focuses on fairness investigations and Section 6 concludes with discussion.

## 2. Existing sample size approach to precisely estimate overall risk and minimise overfitting for a time-to-event outcome

Our previously proposed approach calculates the minimum required sample size for prediction model development,[13, 19] to meet the following criteria:

- criterion (i): a precise estimate of the overall outcome risk by a particular time-point (e.g., absolute margin of error ≤ 0.05)
- criterion (ii): small overfitting of predictor effects defined by an overall shrinkage factor ≥ 0.9.
- criterion (iii): small optimism in apparent model fit defined by Nagelkerke R-squared ($R^2_{Nagelkerke}$), such as ≤ 0.05

Details of the calculations are provided in our previous papers.[13, 19] The approach is implemented in the Stata or R module *pmsampsize*,[14, 15] with the user needing to specify the overall outcome risk (prevalence) at the time-point of interest, the anticipated model performance in the target population (quantified by $R^2_{Nagelkerke}$, Cox-Snell R-squared ($R^2_{CS}$), or the C-index), and the number of candidate predictor parameters for the model. For example, consider developing a model to estimate the 5-year recurrence risk of breast cancer, based on a core set of six predictor parameters and assuming a $R^2_{CS}$ of 0.14 (corresponding to a C-index of 0.70), with a mean follow-up of 3.57 years and event rate of 0.099 recurrences per person year. Applying our previous approach via *pmsampsize* suggests a minimum of 355 participants are required for model development, with about 126 recurrences during 1268 person-years of follow-up.

## 3. Methods: a new sample size approach to target precise estimates of individual risk and improve model fairness

We now introduce a six-step approach to examine the sample size required for model development to target sufficiently precise individual risk estimates at a particular time-point of interest. The process requires specifying the joint distribution of a core set of predictors and an assumed exponential survival model. A closed-form decomposition of Fisher's unit information matrix is then used to derive anticipated uncertainty intervals for a particular sample size. This is implemented using our accompanying *pmstabilityss* module in Stata (see code at https://github.com/JoieEnsor/pmstabilityss-TTE), with R and Python versions forthcoming.

### 3.1 Step-by-step guide to the proposal sample size calculation

**Step (1) - identify a core set of predictors, variables linked to fairness, and a key time-point of interest:**

The precision of risk estimates at the individual level depends on the (joint) distribution of predictors (features) that will be included in the developed model. Therefore, to inform the sample size calculation, the first step is to identify a set of *core predictors*; that is, clinical variables well-known to contribute important key predictive information. Core predictors (e.g., age and stage of disease for cancer outcome prediction) can be identified from previously published models, systematic reviews of prognostic factor studies[20] and conversations with clinical experts. Though additional potential predictors might be considered during the actual model development, the premise is that these additional variables typical add little or no predictive information beyond the core predictors, and may even increase instability and imprecision. In other words, the set of core predictors represent a fundamental basis for targeting precise individual-level risks and define the bare minimum for which we want to control (in)stability for.

Further to the core predictors, it may be important to also identify variables linked to fairness checks. For example, regardless of whether they are included in the developed model, variables such as age, sex, ethnicity, and others (e.g., those representing protected characteristics and subgroups) might be identified so that ultimately (see step (6)) the researcher can examine the sample size required to target sufficiently precise predictions for each subgroup.

Step (1) should also identify a key time-point of interest for prediction; that is, the time-point for which the model will estimate the risk of having the outcome event by. If multiple time-points are relevant, then the sample size calculation need to be repeated for each time-point, as the precision of a risk estimate depends on the time-point chosen. We focus on one key time-point here onwards.

**Step (2) – specify the joint distribution of the core predictors:**
The joint distribution of predictors that are included in the final prediction model impacts the standard errors of the model parameter estimates (e.g., from a Cox model or a parametric survival model), and influences the width of uncertainty intervals around individual-level risk estimates. Hence, step (2) requires the user to specify the joint distribution of core predictors selected in step (1). How to achieve this will depend on the availability of the model development dataset, as follows:

- ***Dataset is already available for model developers:*** this is the most common and simple scenario, as the joint distributions of predictors (and event/censoring times – see step (4)) are already observed. Thus, the user does not need to do anything in this step, as this existing dataset can be used directly in subsequent steps where needed (e.g., in step (5) to derive the unit information matrix).
- ***Dataset exists but not yet available for model developers*** **(e.g., access to a previously collected dataset is conditional on funding success)**: the data holders could be contacted and asked to provide a synthetic dataset that mimics the joint predictor and event (censoring) time distributions, for instance as obtained by a simulation-based approach that models conditional relationships,[21] using packages such as *synthpop* in R.[22] Section 4.2 showcases this approach, but another notable example is the Clinical Practice Research Datalink (CPRD), who have generated synthetic datasets to aid researchers improve workflows (https://www.cprd.com/synthetic-data). Alternatively, data holders could provide summary details of the joint distributions (e.g., cross-tabulations of categorical variables; variance-covariance matrix of continuous variables), to allow the user themselves to simulate a large synthetic dataset containing predictor values for, say, 10,000 individuals. Previous studies that already use the dataset of interest may have summarised baseline variables (e.g., means, standard deviations (SDs), proportions) in their articles, for example within baseline characteristics table, though without covariances.

- ***New data collection required (e.g., a planned prospective cohort study):*** when new data are required, the joint predictor distribution could be based on summary information (e.g., means and SDs for continuous predictors; proportions in each group for categorical predictors) from previous studies in the same target population, for example using published tables of baseline characteristics. These can then be used to simulate a large synthetic dataset of predictor values, as detailed above. A challenge is that often only the marginal distribution for each predictor will be summarised (e.g., from published tables of baseline characteristics), and the correlation or conditional relationship amongst predictors will be unknown. In this situation, a pragmatic starting point is to assume the predictors are conditionally independent, but the impact of this should be examined through sensitivity analyses.

**Step (3) - specify a 'core model' for how individual risks depend on core predictor values:** Alongside an individual's predictor values, the uncertainty around an individual's risk estimate at a particular time-point depends on their underlying event rate as a function of time. Therefore, the user should specify a model that expresses how an individual's event rate over time depends on the values of core predictors from step (2) (referred to as the 'core model'). For example, a parametric time-to-event regression model could be specified, such that the log event rate is a function defined by a baseline log hazard function and beta (log-hazard ratio) terms, which are chosen to match the overall risk at the time-point of interest (e.g., from a Kaplan-Meier curve) and core predictor effects from previous studies in the same population. Often, the 'core model' will be based on an existing model where the aim is to extend a previously published model (e.g., by adding more predictors), or update all parameters of an existing model (e.g., due to concerns of calibration drifts affecting intercept and predictor effect estimates).

The simplest, and thus most practical approach, is to assume the 'core model' is an exponential proportional hazards regression model, such that each participant $i$ has a constant hazard rate of $\eta_i$ and event times ($t_i$) are exponentially distributed conditional on values of P core predictors. This model can be written as:

$$t_i \sim \text{exponential}(\eta_i)$$
$$\ln(\eta_i) = \mu_i = \alpha + \beta_1 x_{1i} + \beta_2 x_{2i} + \cdots + \beta_P x_{Pi}$$
Eq.(1)

Apart from the intercept $\alpha$ (which defines the constant baseline hazard), all parameters correspond to log rate (hazard) ratios, quantifying the change in log rate for a 1-unit increase in the corresponding $x$ value. Eq.(1) can equivalently be written as an accelerated failure time (AFT) model, as follows,[23]

$$\ln(t_i) = -\mu_i + \varepsilon_i = -\alpha - \beta_1 x_{1i} - \beta_2 x_{2i} - \cdots - \beta_P x_{Pi} + \varepsilon_i \qquad \text{Eq.(2)}$$

where $\varepsilon_i$ follows an extreme value distribution with probability density function (pdf) $f(\varepsilon_i) = \exp(\varepsilon_i)\exp(-\exp(\varepsilon_i))$. We will use this AFT specification when deriving the variance matrix of parameter estimates in step (5), but the interpretation of parameters is identical to before, and each patient contributes data $\boldsymbol{x_i} = (1, x_{1i}, x_{2i}, \ldots, x_{Pi})$. Censoring is considered in steps (4) and (5).

### *Rationale for choosing an exponential distribution*

The exponential model is a pragmatic choice: assuming a constant baseline hazard rate ($\alpha$) makes the 'core model' specification (and thus the whole sample size calculation) easiest for researchers. In many situations, a non-constant (potentially even non-linear) baseline hazard rate might be preferable, for example as defined by a parametric distribution such as a Weibull, log-normal or gamma distribution. However, this ultimately creates complexity for providing analytic sample size solutions for the uncertainty of risk estimates, because the more complex parametric distributions require extra (ancillary) parameters than those for the exponential model (e.g., shape and scale parameters for a Weibull model). The uncertainty of these ancillary parameters needs to be accounted for when deriving uncertainty intervals for prediction, which is challenging as predictions depend on a complex (non-linear) function of $\mu_i$ and the ancillary parameters. Although this is achievable post model fitting, here we are focused on sample size calculations before any data analysis and so use the exponential model, as it provides analytic solutions that are closed form and disentangle sample size from other aspects (see step (6)). We note that, for the same reason, analytic (i.e., not simulation-based) sample size calculations for survival data in other settings (e.g., randomised trials evaluating treatment effect) often assume exponential and proportional hazards for simplicity to produce closed-form solutions. In Section 6, we empirically evaluate the impact of assuming the exponential distribution compared to using Weibull or flexible parametric models.

### *Approaches for specifying the 'core model'*

A 'core model' may be difficult to specify; for example, predictor effects may not be readily available if previous models were incompletely reported. We address this by outlining two approaches to simplify the process, utilising the anticipated C-index of the model and overall risk (for the time-point of interest, defined in step (1)) in the target population, alongside assumptions of relative predictor effects (weights).

- ***Approach (a): Specify the overall risk at a particular time-point, C-index, and relative weights of core predictors.*** With this information, we can use an iterative approach to identify an exponential regression equation forming a 'core model' consistent with the specified overall risk and C-index, whilst retaining the *relative* weight of core predictors. This approach simulates predictor values for a large number of participants (based on the distributions specified in step (2)) and then uses an iterative process to identify values of the intercept ($\alpha$) and a multiplicative factor ($\delta$) of the following model,

$$t_i \sim \text{exponential}(\eta_i)$$
$$\ln(\eta_i) = \mu_i = \alpha + \delta(\beta_1 x_{1i} + \beta_2 x_{2i} + \cdots + \beta_P x_{Pi})$$

Eq.(3)

where beta coefficients are the user specified relative weights. Convergence is achieved when the model reaches the specified C-index and overall risk within a small margin of error. Often, Harrell's C-index is known, but other time-dependent C statistics could also be matched as needed.

To simplify the specification of relative weights, we can re-scale variables so that relative effects are the same. Specifically, continuous variables could be re-scaled (e.g., age specified in 10 years) so the effect (log-hazard ratio) of a 1-unit increase (e.g., 10-year increase) is considered to have the same predictive effect as a 1-unit increase in other variables (e.g., smoker versus non-smoker). More broadly, standardisation could be used, as follows in approach (b).

- **Approach (b): specify the overall risk and C-index, whilst assuming same weight of predictors on their standardised scale.** This approach standardises each core predictor (i.e., uses $(x_{1i} - \bar{x}_{1i})/\text{SD}_{x_{1i}}$, $(x_{2i} - \bar{x}_{2i})/\text{SD}_{x_{2i}}$, etc.) from step (2), then the 'core model' of Eq.(3) is specified in terms of these standardised predictors and sets all beta weights to be equal (e.g., = 1). Essentially, this approach assumes the predictive effect of a 1-SD increase to be the same for each standardised predictor. As in approach (a), the iterative approach is then used to identify $\alpha$ and $\delta$ that ensures the 'core model' has the specified C-index and overall risk.

Recall that core predictors in step (2) may include protected characteristics, even if they do not have any predictive effect. If the latter, their relative weight can be set to zero in the 'core model'. Nevertheless, having them in the synthetic or existing dataset ultimately allows prediction and classification uncertainty to be assessed for them.

**Step (4) – derive outcome event times and censoring times:**
As the variance of predictions depends on observed event and censoring times (see step (5)), they need to be available alongside the predictor values. If an existing (full or pilot) or synthetic dataset is available (see step (2)), the outcome event or censoring time can be observed directly for each individual. Otherwise, step (4) requires outcome event times to be generated in the large dataset conditional on the assumed 'core model', for example using the *simsurv* package in R or Stata,[24, 25] which allows the user to specify the survival distribution (e.g., exponential) and predictor effects. Similarly, unless existing or synthetic data are available, the censoring times need to be generated from an assumed survival model of the censoring rate, which usually will be assumed independent of predictors and outcome risk for simplicity. Without other knowledge, it is pragmatic to assume an exponential distribution with a constant censoring rate. Lastly, for each individual define $y_i$ as the minimum of their observed log event time and log censoring time (of course, when using a real dataset, only one of these is observed anyway).

**Step (5) - derive Fisher's unit information after decomposing Fisher's information matrix:**
Steps (5) and (6) involve approximating (based on the information from steps (1) to (4)) the anticipated variance-covariance matrix (var($\widehat{\boldsymbol{\beta}}$)) of model parameter estimates ($\widehat{\boldsymbol{\beta}} = (\hat{\alpha}, \hat{\beta}_1, \hat{\beta}_2, \ldots, \hat{\beta}_P)'$) if we were to fit the assumed 'core model' for a particular sample size. This is needed, as the variance-covariance matrix of the parameters in the 'core model' is subsequently used to calculate the variance of individual-level risk estimates at the time-point of interest.

The key mathematical foundation for Step (5) is to recognise that var($\widehat{\boldsymbol{\beta}}$) (the inverse of Fisher's information matrix) can be decomposed into the total sample size ($n$) and Fisher's *unit* information matrix (**I**):

$$\text{var}(\widehat{\boldsymbol{\beta}}) = n^{-1}\mathbf{I}^{-1} \qquad \text{Eq.(4)}$$

The unit information matrix is defined by,

$$\begin{aligned}\mathbf{I} &= E(\mathbf{XX}' \exp(y_i + \hat{\mu}_i)) \\ &= E(\mathbf{XX}'w_i) \qquad \text{Eq.(5)} \\ &= E(\mathbf{A})\end{aligned}$$

where **X** is the design matrix for the assumed 'core model', with each individual's data corresponding to $\boldsymbol{x_i} = (1, x_{1i}, x_{2i}, \ldots, x_{Pi})$. The $E(\mathbf{A})$ is the expected value of matrix **A** and depends on the joint distribution of the predictors (as this defines **X**), the parameter values of the 'core model' (as this defines $\hat{\mu}_i$) and the observed log follow-up time ($y_i$) for each participant (i.e., the minimum of their log event time and log censoring times).[26] The latter adds complication due to the censoring inherent within time-to-event data (hence Step (4)).

A simple way to derive $E(\mathbf{A})$ is to calculate each of the components of $\mathbf{A}$ for each participant in the (existing or simulated) dataset from step (2), using each participant's observed predictor values and observed event or censoring time ($y_i$), combined with the (exponential) regression parameters of the 'core model' from step (3); the means (across all participants) of each component then provide their expected values and form $\mathbf{I}$. Our module *pmstabilityss* module implements this.

For example, assuming three core predictors and specifying our 'core model' as the following exponential regression,

$$t_i \sim \text{exponential}(\eta_i)$$
$$\ln(\eta_i) = \mu_i = \alpha + \beta_1 x_{1i} + \beta_2 x_{2i} + \beta_3 x_{3i}$$

then $\mathbf{XX}'$ is a 4 by 4 matrix due to the four parameters in the regression equation ($\alpha, \beta_1, \beta_2, \beta_3$), where $\mathbf{X}$ is the design matrix for the assumed 'core model', with each individual's data corresponding to $x_i = (1, x_{1i}, x_{2i}, x_{3i})$. This implies that,

$$\mathbf{I} = E(\mathbf{XX}' \exp(y_i + \hat{\mu}_i)) = E(\mathbf{XX}' w_i)$$

$$= E_{(x,w)} \begin{bmatrix} w_i & x_{1i}w_i & x_{2i}w_i & x_{3i}w_i \\ x_{1i}w_i & x_{1i}^2 w_i & x_{1i}x_{2i}w_i & x_{1i}x_{3i}w_i \\ x_{2i}w_i & x_{1i}x_{2i}w_i & x_{2i}^2 w_i & x_{2i}x_{3i}w_i \\ x_{3i}w_i & x_{1i}x_{3i}w_i & x_{2i}x_{3i}w_i & x_{3i}^2 w_i \end{bmatrix}$$

where $w_i = \exp(y_i + \hat{\mu}_i)$, $\hat{\mu}_i$ is an individual's estimated linear predictor value from the fitted exponential regression (e.g., $\hat{\mu}_i = \hat{\alpha}_i + \hat{\beta}_1 x_{1i} + \hat{\beta}_2 x_{2i} + \hat{\beta}_3 x_{3i}$), and $y_i$ is an individual's log follow-up time (i.e., minimum of the log event time and log censoring time for each patient). To derive the anticipated $\mathbf{I}$ (automated in *pmstabilityss*):

- Set the parameter values in the 'core model' (i.e., $\alpha$, $\delta$, and all $\beta$s) at their anticipated true values, defined in step (3)
- For each individual in the existing or synthetic dataset, define their $y_i$ value following step (4)

- set each individual's $x_{1i}$, $x_{2i}$, and $x_{3i}$ values as the (standardised or unstandardised) observed values of core predictors in the (existing or synthetic) dataset from step (2)
- use this information to derive $\mathbf{X}$ and $\exp(y_i + \hat{\mu}_i)$
- derive $E(\mathbf{A})$ by calculating the 16 components of $\mathbf{A}$ for each participant in the (existing or synthetic) dataset, then the mean of each component provides their expected values and thus forms $\mathbf{I}$.

**Step (6) - examine the impact of sample size on precision of individual risk estimates:**

The final step is to examine how sample size impacts the level of precision (uncertainty distribution and interval widths) around individual risk estimates at the key time-point of interest (as chosen in step (1)). This is relevant when the user has an existing dataset (to check if this sample size is large enough) or when designing a new study with prospective data collection (to *a priori* identify the required sample size). These situations are now outlined as Options A and B, below.

- **Option A: Calculate expected uncertainty of predictions for a given sample size (existing dataset)**

Following maximum likelihood estimation of an exponential regression model, the variance of a new individual's log (hazard) rate ($\hat{\mu}_{new}$) is,

$$\text{var}(\hat{\mu}_{new}) = \text{var}(\boldsymbol{x}_{new}\widehat{\boldsymbol{\beta}}) = \boldsymbol{x}_{new}\, \text{var}(\widehat{\boldsymbol{\beta}})\, \boldsymbol{x}'_{new} \quad\quad \text{Eq.(6)}$$

where $\boldsymbol{x}_{new} = (1, x_{1new}, x_{2new}, \ldots, x_{Pnew})$ are the predictor values for the new individual. Substituting in Eq.(4), this can be rewritten as:

$$\text{var}(\hat{\mu}_{new}) = n^{-1}\, \boldsymbol{x}_{new}\, \mathbf{I}^{-1}\, \boldsymbol{x}'_{new} \quad\quad \text{Eq.(7)}$$

Hence, for a specified sample size ($n$), we can derive an anticipated 95% uncertainty interval around an individual's log-rate using,

$$\left[\hat{\mu}_{new} \pm \left(1.96 \times \sqrt{\text{var}(\hat{\mu}_{new})}\right)\right] = \left[\hat{\mu}_{new\_L}, \hat{\mu}_{new\_U}\right] \quad\quad \text{Eq.(8)}$$

where $L$ and $U$ denote lower and upper, respectively. However, the actual scale of interest is the survival probability scale ($S(t)$), or the event risk scale ($F(t) = 1 - S(t)$). Focusing on the latter, we can derive an individual's true event risk (based on the 'core model') at a particular time-point ($t$) using

$$F_{new}(t) = 1 - \left(\exp(-\exp(x_{new}\widehat{\boldsymbol{\beta}}))t\right) \qquad \text{Eq.(9)}$$

with $\widehat{\boldsymbol{\beta}}$ replaced by the true parameters of the core model. Then, the anticipated 95% uncertainty interval for an individual's event risk at $t$ is given by:

$$\left[1 - \exp(-\exp(\hat{\mu}_{new\_L}))t, \quad 1 - \exp(-\exp(\hat{\mu}_{new\_U}))t\right] \qquad \text{Eq.(10)}$$

Therefore, Option A requires the user to calculate Eq.(7) to Eq.(10) and derive uncertainty intervals for the risk of each individual from the target population, conditional on a specified model development sample size ($n$) and a time-point ($t$) of interest. The time-point was already chosen in step (1); the unit information matrix (**I**) from Step (5); and the new participants can just be those from the existing or simulated dataset from Step (2) which already contain predictor values ($x_{new}$). Thus, to apply Eq.(7) to Eq.(10) all that remains is to specify the sample size of interest; this could be the available number of participants in the existing dataset being considered for model development, or it might be a specified sample size being considered for new data collection (e.g., that determined by *pmsampsize*). Sometimes, a range of different sample sizes might be considered to examine the value of information (e.g., in terms of reduced width of uncertainty intervals, reduced classification instability) arising from including additional participants over and above that recommended by *pmsampsize*.

- **Option B: Calculate a target sample size for new data collection to ensure precise individual-level predictions**

When designing a new study to recruit participants for model development, researchers will need to calculate the sample size required to target particular precision of risk estimates. By rearranging Eq.(7), the sample size needed to target a chosen variance of the log-rate estimate for an individual is:

$$n = \text{var}(\hat{\mu}_{new})^{-1} x_{new} \mathbf{I}^{-1} x'_{new} \qquad \text{Eq.(11)}$$

Eq.(11) can then be applied to each individual in the (real or simulated) dataset from Step (2) to obtain the required $n$ for their particular combination of predictor values ($x_{new}$). A practical issue is how to select the target value of var($\hat{\mu}_{new}$) for each individual, as this is on a difficult scale to interpret. Further, the required value of var($\hat{\mu}_{new}$) will not be consistent across individuals, due to the multiplication with $x_{new}\ \mathbf{I}^{-1}\ x'_{new}$, which is individual specific. A pragmatic approach is to specify the maximum var($\hat{\mu}_{new}$) allowed for a range of $F_{new}(t)$ values (e.g., 0.01, 0.025, 0.05, 0.10, 0.15, 0.20, etc), corresponding to a target maximum uncertainty interval width on the event risk (i.e., $F_{new}(t)$) scale via Eq.(9) and Eq.(10)). These can then be applied to each individual by using the var($\hat{\mu}_{new}$) value that corresponds to the categorised $F_{new}(t)$ value closest to their true $F_{new}(t)$. Special attention may also be given to selecting appropriate var($\hat{\mu}_{new}$) values in particular subgroups defined by combinations of predictor values (e.g., sex, ethnicity), where algorithmic fairness checks will be important.

**3.2 Deciding and presenting target uncertainty intervals with patients and clinical stakeholders: perspective based on risk thresholds and decision theory**

Regardless of whether option A or B is chosen, model developers will need to decide what width of uncertainty intervals they deem appropriate. Ideally, a suitably narrow interval is desired for *every* individual (for all combinations of values of predictors in the 'core model'). However, depending on the clinical context and role of the model for clinical practice, some regions of estimated risk may not require intervals to be as narrow as in other regions. For instance, having wide uncertainty intervals for individuals with high risk (e.g., reflected by uncertainty intervals from 0.3 to 0.95) may not matter if the entire interval range is compatible with a perceived high risk. This concept aligns with preferences for risk thresholds for clinical decisions. For example, a 10-year CVD risk threshold of 0.1 (i.e., 10%) is typically used to guide decisions to prescribe statins, and so wide uncertainty intervals that span 0.3 to 0.95 might be deemed acceptable, but narrower intervals that span 0.05 to 0.3 (including the threshold of 0.1) may not. When deciding on appropriate uncertainty interval widths, it is important to understand the clinical context of how the model will be used to guide decision making and any corresponding risk threshold(s) involved.

With this in mind, it will be helpful to identify risk thresholds within a decision-theory perspective,[27-29] based on preferences (utilities) elicited from patients, clinicians and other relevant stakeholders about particular outcomes and consequences that may follow from possible decisions. For example, a VTE 3-year recurrence risk threshold of about 5% is suggested for when an individual should be recommended to continue anti-coagulation treatment.[1] If a well-calibrated prediction model estimates the individual's risk to be ≥ 5%, then this suggests the correct decision is to continue anti-coagulation treatment. However, there may still be uncertainty about the suggested decision due to uncertainty of the model's risk estimate; if the uncertainty is too wide, then ideally additional information is needed before making a decision.[30]

In this context, the aim of our sample size approach is to help understand and examine which sample sizes are likely to give sufficient information to guide decisions at the individual-level. To help examine this, we recommend calculating and presenting to stakeholders:

- *Prediction instability plots*, where each individual's 'true' risk from Step 3 (x-axis) is plotted against their corresponding uncertainty interval (y-axis) from Step 5. The question to ask stakeholders is whether the individual uncertainty intervals are too wide for using or endorsing the model in practice. To facilitate this discussion, we recommend prediction instability plots are presented with two curves (e.g., using a LOWESS smoother or spline function) fitted separately through individuals' upper and lower uncertainty interval values. These curves define a 'typical' 95% uncertainty interval at each risk, across the entire spectrum of estimated risks from 0 to 1, which should aid visual interpretation for stakeholders (as individual uncertainty intervals can vary considerably, even for those with the same estimated risk). This is demonstrated in Section 4.2.
- *Classification instability plots (if risk thresholds are relevant)*, plotting each individual's 'true' risk (x-axis) (i.e., the risk defined by the 'core model') against the proportion (y-axis) of their uncertainty distribution that falls on the opposite side of their chosen clinical risk threshold compared to their 'true' risk. The question to ask stakeholders is whether, in general, the proportion of the uncertainty distribution on the opposite side of the threshold is generally too large for them to use or endorse the model for individuals in practice. In our examples in Section 4, we assume the same risk threshold is relevant for all individuals, but this can be relaxed if stakeholders recommend different thresholds across particular subgroups.

- *Summary statistics,* that quantify the magnitude of uncertainty and classification instability across individuals. In particular, the mean (min, max) width of 95% confidence intervals, and the mean (min, max) probability of misclassification. Also, the mean (min, max) across individuals of their mean absolute prediction error (MAPE), or root mean squared prediction error (RMSPE), which can be derived by many (e.g., 1000) sampling values from each individual's uncertainty distribution and calculating mean absolute (or root mean-squared) differences to their 'true' risk.
- *Subgroup plots and results,* that summarise the anticipated uncertainty and classification instability in relevant subgroups of people (e.g., defined by sex, ethnicity – see Section 5).

We now illustrate all these ideas with two examples applying our new approach. In the supplementary material S1, we also discuss how to measure the impact of uncertainty on clinical utility, using the net benefit function.[31]

## 4. Results I: Application to a model for breast cancer recurrence by 5 years

Consider the scenario where researchers want to develop a prediction model for estimating the risk of breast cancer recurrence within 5 years in those with node positive and oestrogen-receptor (ER) positive breast cancer treated with tamoxifen. They plan to recruit participants to a new cohort study and want to know the sample size required to produce a reliable model. Using *pmsampsize* and assuming a mean follow-up of 3.57 years and event rate of 0.099 recurrences per person year, a minimum of 355 participants are recommended to identify the sample size needed to target population-level stability in estimated risks. As the model is to be used to guide individual-level counselling and decision making, it is also important to target precise estimation of individual-level risk estimates. This can be done using the six-step process outlined in Section 3 and we now describe this process, under three scenarios. In Section 4.1 we assume that an existing dataset is available to inform the calculations. Sections 4.2 and 4.3 consider that, in the absence of an existing dataset, a synthetic dataset has been generated and provided by external researchers, either with (Section 4.2) or without (Section 4.3) follow-up information included.

**4.1 Sample size calculation based on an existing dataset at hand**

A previous dataset is available of 686 patients diagnosed with node positive breast cancer from 1984 to 1989 and recruited to the German Breast Cancer Study Group (GBSG). The GBSG dataset contains five core predictors considered important to predict 5 year survival (see Step (1)) alongside information about follow-up and any recurrence times.[32, 33] Here, to inform our sample size calculations, we focus on the subset of 220 participants that were ER positive and received tamoxifen treatment, which represents our target population of interest for recruiting patients to develop a new prediction model. Henceforth, we consider this dataset as akin to a pilot study and *pmstabilityss* uses it within the six-step process of Section 3 to examine required sample sizes.

*Step (1) - identify a core set of predictors:*

Five core predictors were identified: age (years), tumour size (mm), number of positive lymph nodes, menopausal status (pre or post) and tumour grade (1, 2 or 3). This corresponds to 6 predictor parameters. To help with specification of relative weights (see Step (3)), the three continuous predictors were standardised (e.g., age was specified as $(\text{age}_i - \overline{\text{age}}_i)/\text{SD}_{\text{age}_i}$).

*Step (2) - specify the joint distribution of the core predictors:*

The joint distribution of the five core predictors was observed in the 220 participants from the GBSG pilot dataset; thus, no synthetic data was needed. Hence, subsequent prediction and instability plots (see Step (6)) are based on these 220 participants.

*Step (3) - specify a 'core model' for how individual risks depend on core predictor values:*

The 'core model' was chosen to be an exponential proportional hazards model with relative predictor weights identified from analysis of the GBSG dataset itself, and for simplicity, assuming linear functional form for continuous predictors on their standardised scale:

$$t_i \sim \text{exponential}(\eta_i)$$

$$\ln(\eta_i) = \mu_i$$

$$= \alpha + \delta((-1 \times \text{age}) + (0.5 \times \text{size}) + (2 \times \text{\#nodes})$$

$$+ (3 \times \text{post\_menopause}) + (3 \times \text{grade2}) + (4 \times \text{grade3}))$$

The $\alpha$ and $\delta$ values were set to be -3.429 and 0.208, respectively, identified based on trial and error, to ensure that the 'core model' had a C-index of 0.70 and a mean estimated risk of 0.39 in the 220 GBSG patients, in order to match the observed performance and overall 5-year recurrence risk in the dataset. Recent survival statistics for ER-positive patients also reported a similar 5-year recurrence risk.[34] For brevity, the impact of changing these predictor weights is considered in supplementary material S3.

*Step (4) - derive outcome event times and censoring times:*

The time of either censoring or recurrence was available for each participant in the GBSG dataset, so allowed us to define their $y_i$ variable (i.e., log follow-up time, representing the minimum of log time of censoring or recurrence). Using the $\hat{\mu}_i$ from Step (3), we calculate $w_i = \exp(y_i + \hat{\mu}_i)$ values toward Fisher's unit information matrix. The mean follow-up time was 3.57 years across all participants; 4.11 years in those that were censored before recurrence; and 2.59 years in those that had a recurrence. The maximum follow-up length was 7.28 years; the total follow-up was 785.4 person-years; and the overall event rate was 0.099 (=78/785.4) recurrences per person year.

*Step (5) - derive Fisher's unit information after decomposing Fisher's information matrix:*

Apply Eq.(5) to derive the unit information matrix (**I**).

*Step (6) - examine the impact of sample size on precision of individual risk estimates:*

We focus on estimates of risk at 5 years and examine how sample size impacts the width of 95% uncertainty intervals around risk estimates. We summarise the probability of misclassification based on an illustrative risk threshold of 20%, for example to identify high-risk individuals for whom chemotherapy might be considered (the impact of using a different threshold is considered in Supplementary material S3). We begin by considering the sample size recommended by *pmsampsize*, which was 355 participants (see Section 2). Prediction and classification instability plots are shown in Figure 1(a), and summary statistics in Table 1.

The anticipated 95% uncertainty intervals around individual risk are quite narrow, with a mean width of 0.23 and a mean MAPE of 0.046. The probability of misclassification (proportion of the uncertainty distribution below/above the threshold value of 20% when 'true' risk is actually above/below) is generally very low, with a mean probability of 0.057. Those with very high 'true' risks have a MAPE of about zero. However, MAPE is higher for individuals with 'true' risks ≤ 0.3 and some have misclassification probabilities over 0.1, even toward 0.5, with a mean of 0.18. Therefore, a sample size of 355 participants might be considered too low, if greater precision of risk estimates is needed for those with 'true' risks ≤ 0.3, especially as (e.g., chemotherapy) decisions for this group are very sensitive to understanding their actual risk.

This issue could be discussed via Patient and Public Involvement and Engagement (PPIE) groups and with clinical stakeholders, to identify target precision required, which may differ across the spectrum of 'true' risks from 0 to 1. Then, we can apply Eq.(11) to identify the sample size required to target the precision required. For example, consider that stakeholders recommend targeting 95% uncertainty interval widths < 0.20 for all individuals with a 'true' risk of ≤ 0.3 (Figure 1(b)). Applying Eq.(11) identifies a total sample size of about 920 participants is needed to achieve this; this sample size would also reduce this subgroup's anticipated mean misclassification probability down from 0.18 to 0.12. Note, the more stringent the precision and misclassification criteria required, the larger the sample size needed.

**Figure 1** Prediction and classification instability plots for the breast cancer recurrence model assuming a particular sample size, using the GBSG dataset of 220 patients for informing the joint distribution of predictors, censoring and recurrence times (dash lines correspond to 95% uncertainty intervals). Classification is based on a risk threshold of 20%.

(a) 355 participants (minimum recommended by *pmsampsize*)

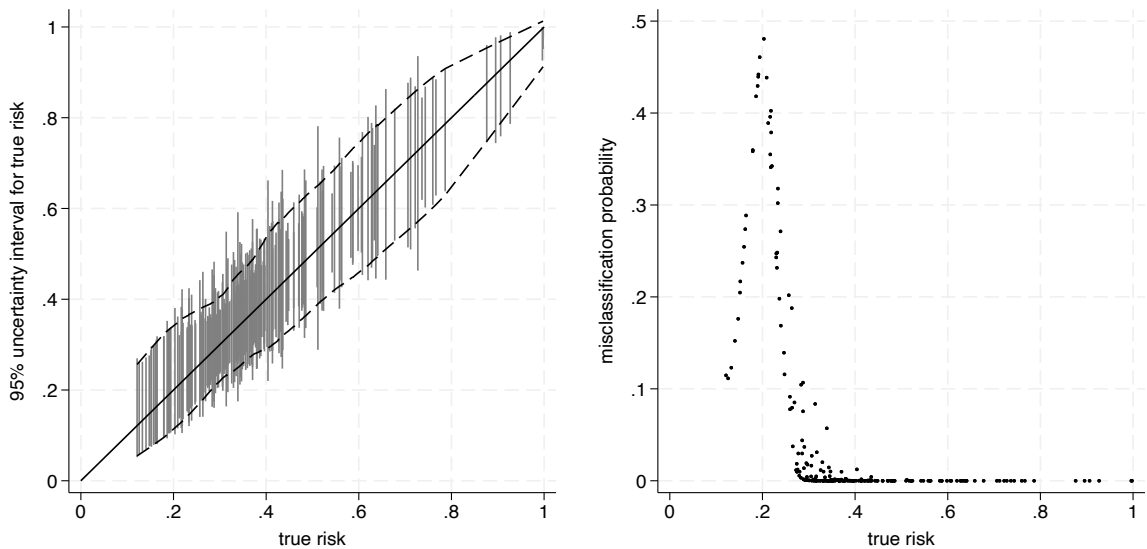

(b) 920 participants (to target 95% uncertainty interval widths ≤ 0.2 in those with true risks ≤ 0.3)

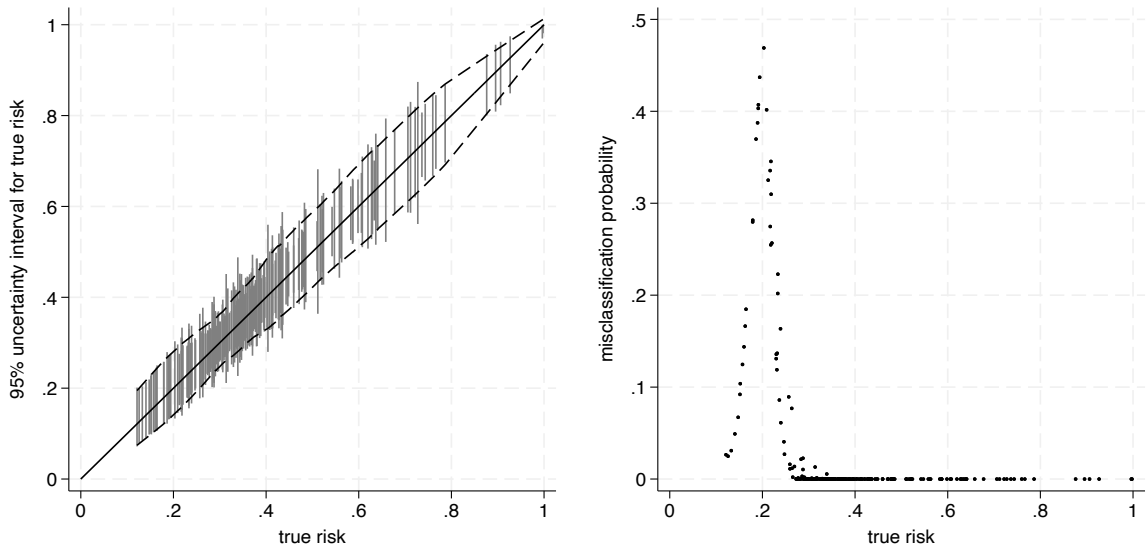

*Table 1: Summary statistics for the expected precision and probability of misclassification for the breast cancer model if developed using particular sample sizes and based on particular assumed core models and available data*

| Data used to inform sample size calculations; assumed core model | Sample size for new model development | 95% uncertainty interval width: *mean (min, med, max)* | Mean Absolute Prediction Error (MAPE): *mean (min, med, max)* | Probability of misclassification based on 20% threshold: *mean (min, med, max)* |
|---|---|---|---|---|
| GBSG dataset of 220 patients; original core model - see Section 4.1 | 355 (minimum no. recommended by *pmsampsize*) | 0.23 (0.048, 0.22, 0.49) | 0.046 (0.0066, 0.044, 0.11) | 0.057 (0, 0.00016, 0.48) |
| | 920 (no. to target 95% uncertainty interval widths ≤ 0.2 in those with true risks ≤ 0.3) | 0.14 (0.019, 0.13, 0.32) | 0.029 (0.0032, 0.027, 0.066) | 0.037 (0, 0, 0.47) |
| Synthetic dataset of 10000 patients provided with predictor values & survival times; original core model - see Section 4.4 | 355 | 0.22 (0, 0.21, 0.56) | 0.046 (0, 0.043, 0.12) | 0.054 (0, 0.000035, 0.50) |
| | 920 | 0.14 (0, 0.13, 0.37) | 0.029 (0, 0.028, 0.084) | 0.037 (0, 0, 0.50) |
| Synthetic dataset of 10000 patients provided with predictor values but without survival times; original core model - see Section 4.5 | 355 | 0.22 (0, 0.21, 0.58) | 0.046 (0, 0.044, 0.12) | 0.055 (0, 0.000046, 0.50) |
| | 920 | 0.14 (0, 0.13, 0.39) | 0.030 (0, 0.028, 0.081) | 0.037 (0, 0, 0.50) |

**4.2 Sample size calculation based on a synthetic dataset including follow-up information**

Now let us consider that the GBSG dataset is not available, for example due to it being held externally and unavailable for transfer (e.g., due to confidentiality constraints or access restrictions). In that situation, researchers could ask the data holders to generate a large synthetic dataset to mirror the real dataset, in terms of the joint distribution of predictors, and censoring and recurrence times. This synthetic dataset could then be used in the sample size calculation. To illustrate this, we used the *synthpop* library in R,[22] to produce a synthetic dataset of 10000 participants with randomly generated values of the six core predictors alongside follow-up time and whether they were censored or had a recurrence at that time. Example code is provided in supplementary material S2. Briefly, this approach utilises the joint distribution of variables as carefully specified in terms of a series of conditional distributions, similar to how multiple imputation is undertaken using chained equations. Simulating a large dataset of 10000 participants ensures (if the conditional distributions are carefully chosen) that the observed (marginal and conditional) distributions of variables in the original GBSG dataset will be closely reflected by the synthetic data (e.g., in terms of means and SDs of continuous variables; category proportions of categorical variables). This can be checked by the original data holders, before finalising their synthetic data to be passed over.

Once the synthetic data are obtained, the researchers can use this to apply the sample size calculation (and *pmstabilityss*) following the same steps as described in Section 4.1. Note that the large size of the synthetic data (10000 participants) is simply to help ensure the distributions of predictors and other variables closely reflect those in the (unseen) GBSG dataset; it is <u>not</u> the sample size required for model development. The synthetic dataset is used to generate the unit information matrix in Step (5), and then Step (6) examines the actual sample size required for model development based on this unit information matrix.

The results based on the synthetic data are summarised in Table 1 and Figure 2; they are very similar to when the GBSG dataset was available directly, with similar summary statistics about uncertainty interval widths and misclassification probabilities. For example, with 355 participants the mean MAPE is 0.046 when using either dataset, and the mean uncertainty interval width is now 0.22 compared to 0.23 before. Hence, similar conclusions about the sample size required would be achieved in general.

*Figure 2 Prediction and classification instability plots for the breast cancer recurrence model assuming a particular sample size, using a synthetic dataset for informing the joint distribution of predictors, censoring and recurrence times*

(a) 355 participants (recommended by *pmsampsize*)

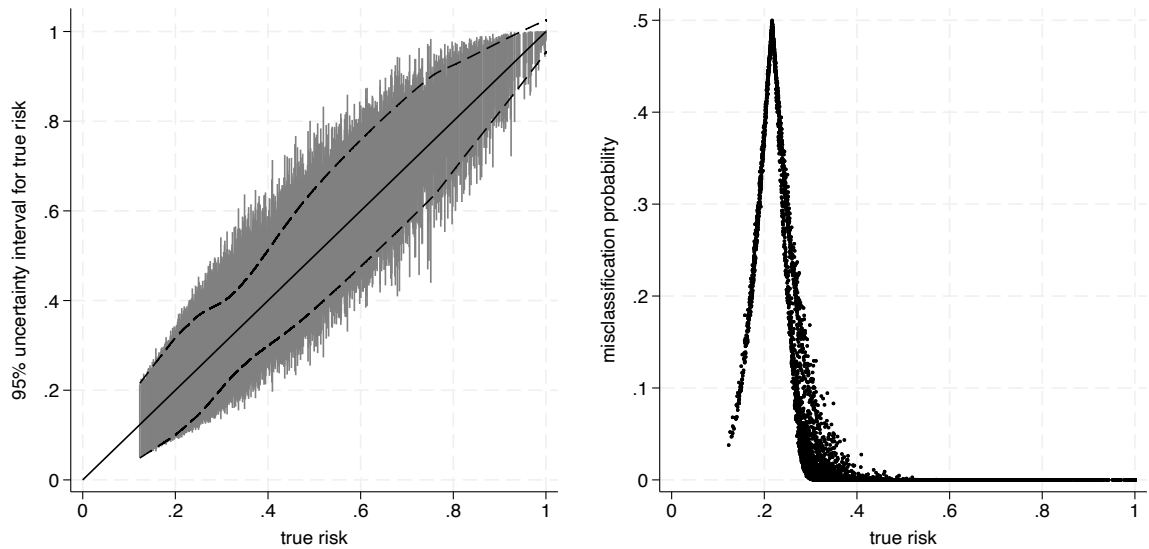

(a) 920 participants

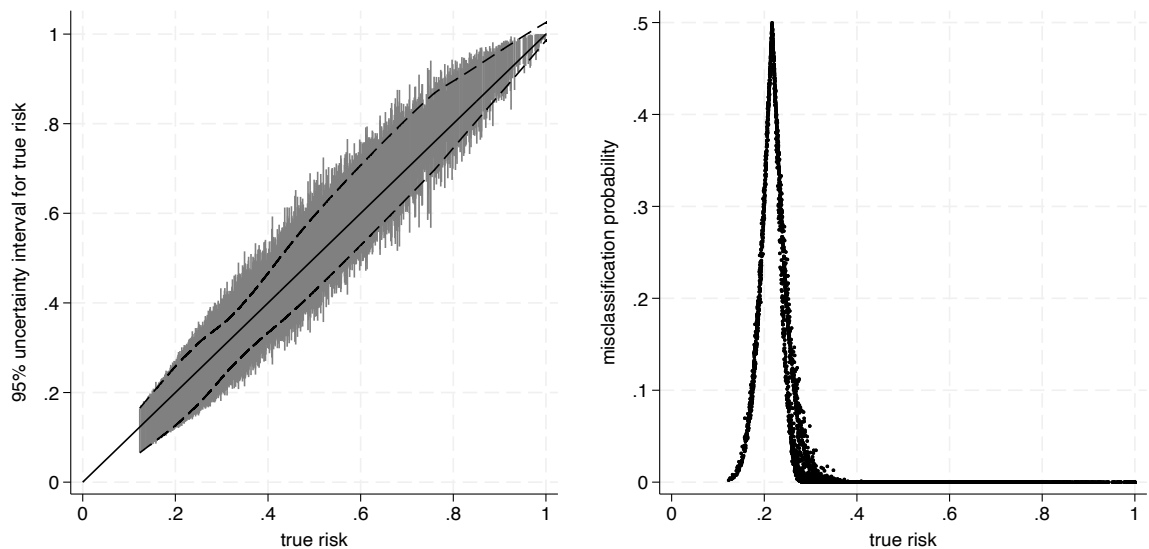

Slight differences do arise in the lower range of predictions, potentially due to the larger size of the synthetic data providing more individuals with lower risks than before. In particular, the mean uncertainty interval width in those with 'true' risks ≤ 0.3 is slightly wider when using the synthetic data than when using the GBSG data directly, leading to a higher mean probability of misclassification (0.22 compared to 0.18 with 355 participants; 0.16 compared to 0.12 with 920 participants). Subsequently larger sample sizes would be recommended based on the synthetic data if targeting uncertainty interval widths ≤ 0.2 in the subgroup with 'true' risks ≤ 0.3.

**4.3 Sample size calculation based on a synthetic dataset without follow-up information**

Sometimes a (synthetic) dataset may be available for informing the joint distribution of predictors but without information about individual-level follow-up times and event status (e.g., due to confidentiality concerns). In this situation, implementing the sample size calculation requires the researcher to generate survival times under some sensible assumptions, including the anticipated maximum follow-up length, the mean follow-up time, and the censoring distribution.

To illustrate this, we revisit the sample size calculation using the synthetic dataset from Section 4.2 but exclude any follow-up information. That is, the synthetic dataset now contains the values of the core predictors for the 10000 participants, so that we have the joint predictor distribution needed for Step (2). However, we do not have the value of $w_i = \exp(y_i + \hat{\mu}_i)$ for each individual needed from Step (4). To address this, each individual's $\hat{\mu}_i$ is set to the $\mu_i$ from the 'core model' from Step (3), and we need to generate $y_i$, which is the minimum of the log event time and log censoring times for each patient. Using *survsim*,[24] we generate the log event time for each patient conditional on their $\mu_i$ defined by the exponential regression of the 'core model'; given the large dataset, this still ensures that the overall risk at 5 years was 0.39, akin to that for the GBSG dataset (example code in Supplementary material S2). Then, we assume no censoring occurs in the first two years (based on the premise that drop out before 2 years is very unlikely), and between 2 and 7.28 years there is uniform censoring, with all remaining individuals censored at 7.28 years (to match the maximum follow-up known for the GBSG). This led to the synthetic dataset having a mean follow-up time of 3.55 years across all participants (recall it was 3.57 years in the GBSG dataset).

The results are summarised in Table 1 and Figure 3, and are very similar to previous examples where follow-up times were directly available either within the GBSG dataset directly or the supplied synthetic dataset. For example, with 920 participants the mean misclassification probability is 0.037, which is the same as previously. When considering just those with true risks ≤0.3, the mean misclassification probability is 0.15 compared to 0.12 when the GBSG dataset was available directly and 0.16 when the provided synthetic data included follow-up times.

*Figure 3 Prediction and classification instability plots for the breast cancer recurrence model assuming a particular sample size, using a synthetic dataset providing the joint distribution of predictors, and then separately generating censoring and recurrence times*

(b) 355 participants (minimum recommended by *pmsampsize*)

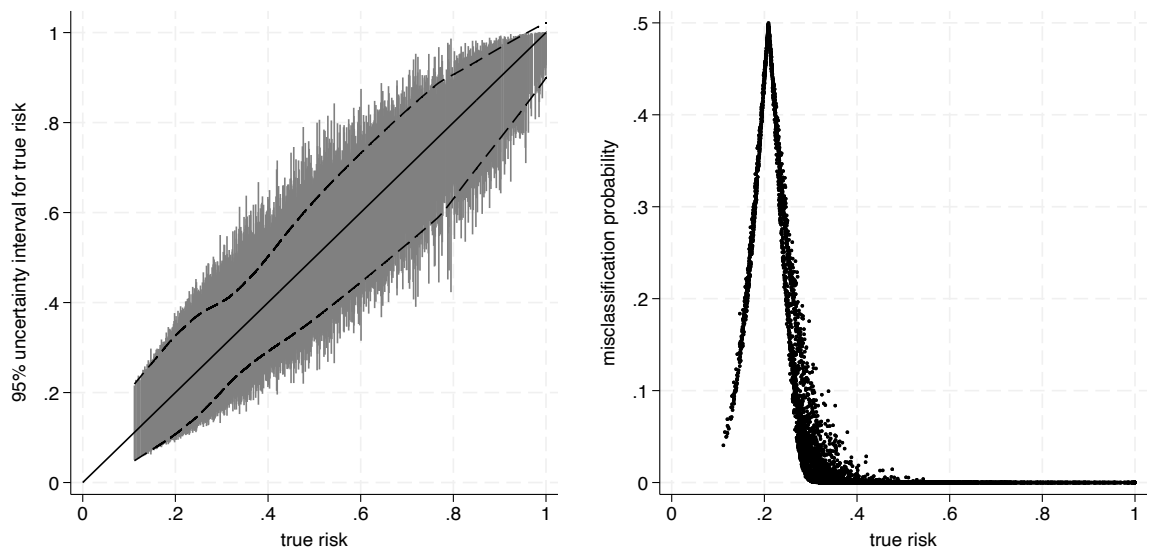

(b) 920 participants

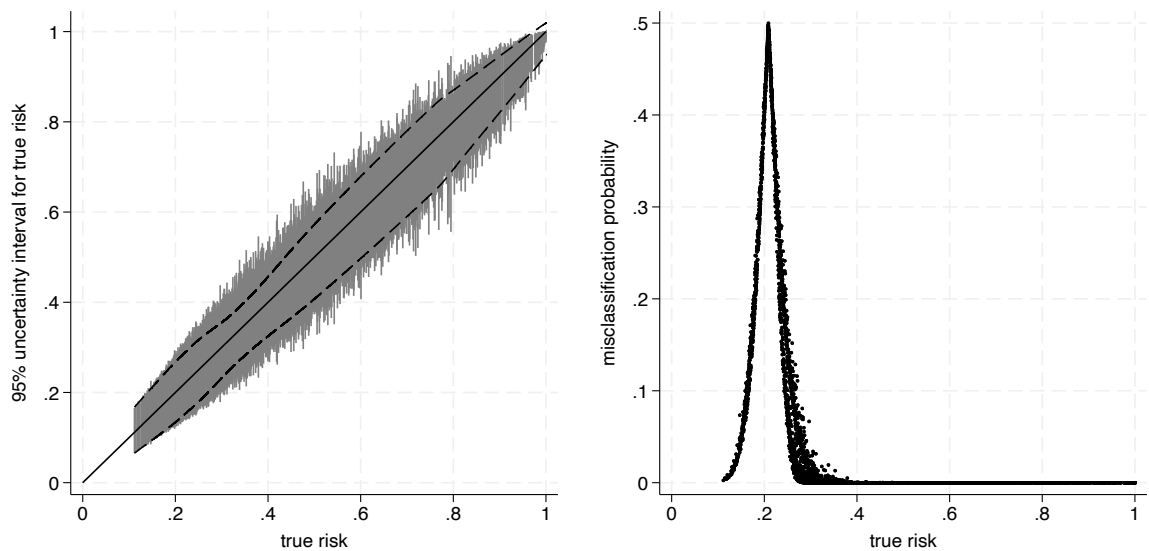

## 5. Results II: Examination of stability in subgroups to inform model fairness checks

Robustness of model predictions in subgroups may form part of fairness checks for using a model in practice, and is a recommendation within the TRIPOD+AI reporting guideline.[35] Thus, as part of the sample size calculations, it may be important to examine anticipated uncertainty intervals and classification instability in subgroups defined by relevant patient characteristics. For this reason, in Step (1) we mentioned that the core predictors may include variables that represent protected characteristics. The model might include such predictors in the 'core model' or leave them out; regardless, precision and classification instability can be checked as long as the relevant variables are available after Step (1).

To illustrate this, we return to the breast cancer example and examine how a sample size of 920 participants is anticipated to impact subgroups defined by menopause status, as information from prediction models should be reliable for women both pre and post menopause. The results (based on having the GBSG dataset available as in Section 4.1) are shown in Table 2 and Figure 4, and suggest that risk estimates will be more uncertain for pre-menopausal women, for example with a mean MAPE of 0.032 compared to 0.027 for post-menopausal women. This is anticipated, as the minority (26%) of participants in the GBSG dataset are pre-menopausal. They also have a higher mean misclassification probability (based on the 20% threshold) of 0.059 compared to 0.030. Nevertheless, the discrepancies between groups are small, and both groups are anticipated to have quite low misclassification probabilities. Thus, we conclude that a sample size of 920 participants is unlikely to provide important fairness concerns in terms of the model's precision and classification ability for different menopausal groups. We recognise that this is just one aspect of fairness.

*Table 2: Summary statistics for the expected precision and probability of misclassification for the AKI model when developed using 920 participants*

| Menopause status | 95% uncertainty interval width: *mean (min, med, max)* | Mean Absolute Prediction Error (MAPE): *mean (min, med, max)* | Probability of misclassification: *mean (min, med, max)* |
|---|---|---|---|
| Pre-menopause | 0.14 (0.019, 0.13, 0.32) | 0.032 (0.003, 0.030, 0.064) | 0.059 (0, 0.00039, 0.37) |
| Post-menopause | 0.13 (0.084, 0.12, 0.20) | 0.027 (0.0050, 0.025, 0.065) | 0.030 (0, 0, 0.47) |

*Figure 4 Prediction and classification instability plots for pre- and post-menopausal women when developing the breast cancer recurrence model assuming a total sample size of 920 participants*

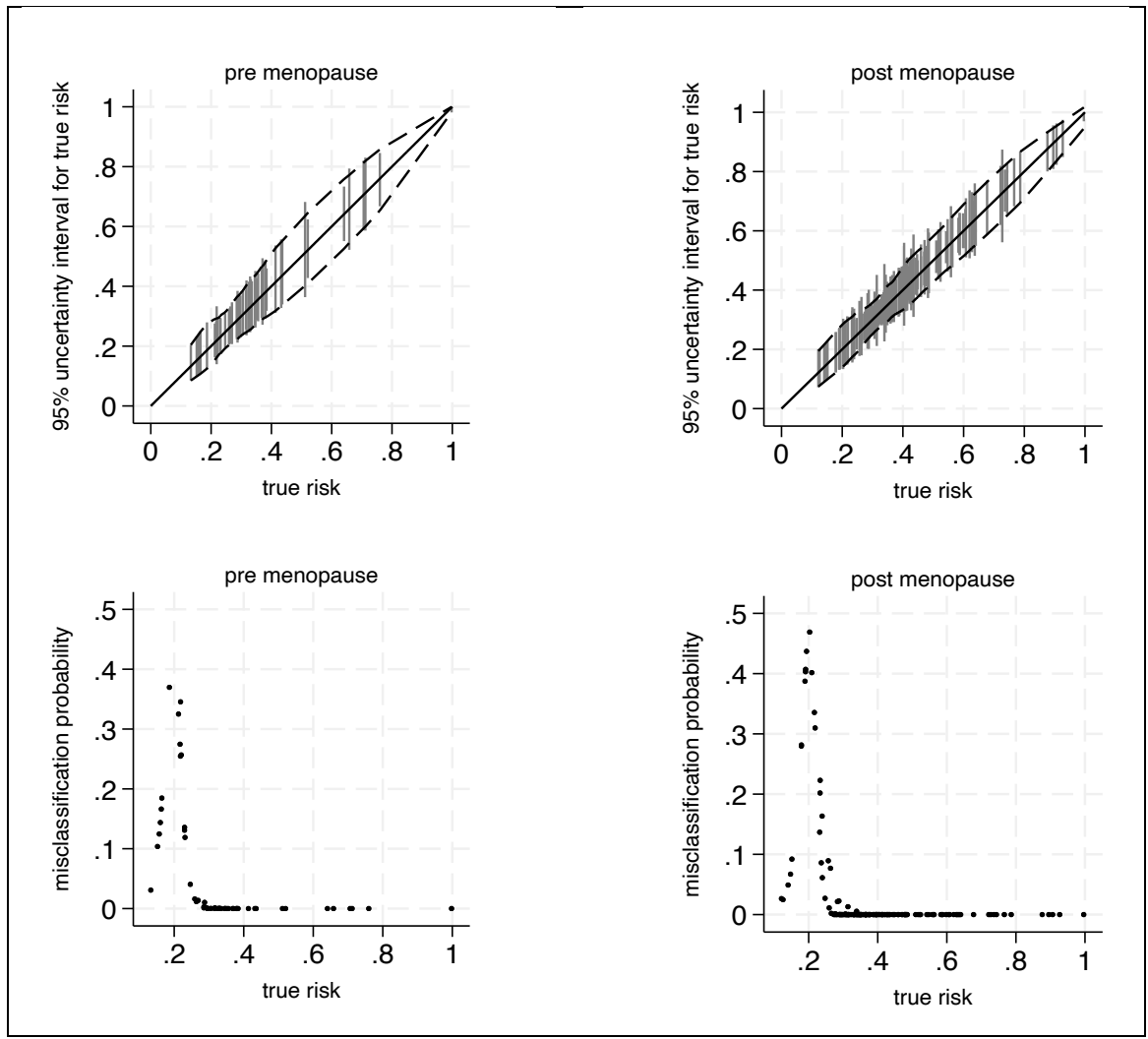

## 6. Results III: Empirical comparison of exponential, Weibull and flexible parametric models

As mentioned in Section 4, specifying an exponential regression as the 'core model' is a pragmatic decision to facilitate closed-form solutions for the sample size calculation. An important question, therefore, is whether these also serve as a reasonable approximation in situations where the baseline hazard is not constant.

To examine this empirically, we fitted exponential, Weibull and flexible parametric regression models (with the same set of predictors as considered in the sample size calculation) to the existing breast cancer dataset of 220 participants, for which the baseline hazard appeared non-constant (Figure 5(a)). The flexible parametric model used restricted cubic splines to model the baseline cumulative hazard,[36, 37] with four internal knots. Post-estimation of each model, we derived 95% uncertainty intervals for the estimated risk of each of the 220 participants; these were calculated by using Eq.(7) to Eq.(10) for the exponential model (i.e., the same approach as in the sample size calculation), and by using the *predictnl* command in Stata for the Weibull and flexible parametric models (which numerically estimates standard errors and confidence intervals directly on the risk scale).

The estimated risks and uncertainty intervals are compared for the various models in Figure 5 and supplementary material S4, and generally quite similar for most participants. For example, the mean interval width is 0.291 for the Weibull model, slightly larger than the mean width of 0.286 for the exponential model. At the individual level, the largest disagreement in interval widths occurs in those with estimated risks close to 1, and in this situation the *predictnl* standard errors and confidence intervals (on the risk scale) are themselves not robust for the Weibull and flexible parametric models, anyway.

We also compared the three approaches when applied to a prostate cancer dataset of 502 participants, to estimate 12-month mortality risk using four predictors. This example also showed a non-constant baseline hazard (Figure 5), but to a smaller degree than the breast cancer example. The risk estimates and uncertainty intervals for the three models were again very similar (Figure 5, supplementary material S5), though the uncertainty intervals were slightly larger from the Weibull and flexible parametric models. We anticipate this will often be the case, due to the extra parameters being estimated.

In summary, these investigations give some reassurance that estimates of prediction uncertainty based on assuming an exponential core model is a good approximation for use in a sample size calculation.

**Figure 5** Baseline hazard rate for breast cancer and prostate cancer datasets, with comparison of risk estimates and their 95% uncertainty intervals when fitting an exponential or flexible parametric survival model

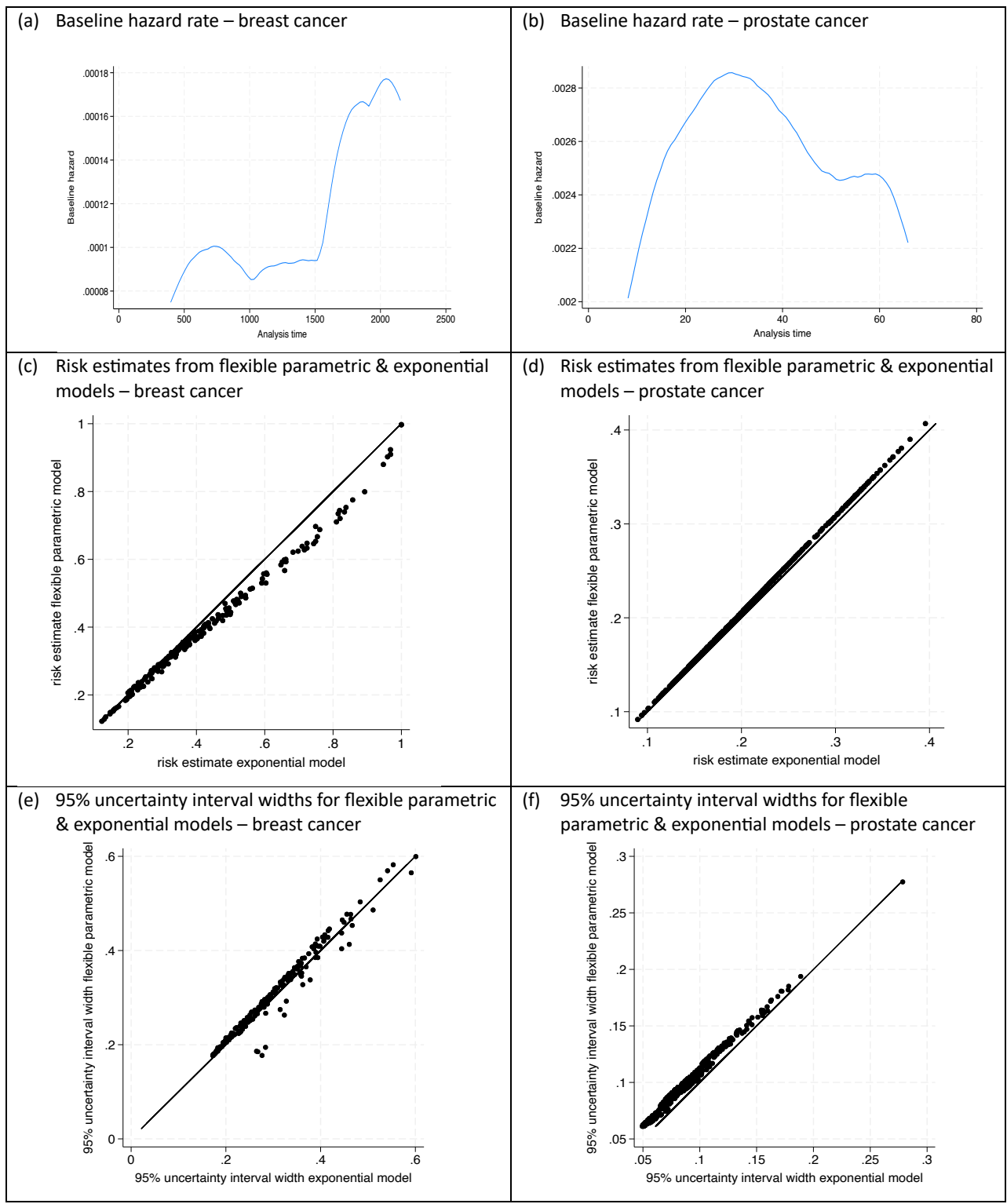

## 6. Discussion

In this article, we have considered the sample size required to develop a clinical prediction model using time-to-event (survival) data from a cohort study. We focused on targeting individual-level stability in risk estimates, building on previous work for binary outcomes.[17] Specifically, we proposed how to examine the (minimum) sample size required to produce individual risk estimates that are precise enough for the clinical context at hand. This can be done to inform the decision to collect new data (e.g., grant application for new model development, or to set-up a new cohort study or extend an existing one) or to help decide if an existing dataset is suitably large for model development. Our software *pmstabilityss* implements the approach and allows users to generate summary plots and statistics. Users might focus on one particular sample size (e.g., as defined by *pmsampsize* or an existing dataset being considered) or examine a range of sample sizes (e.g., for deciding the size of a new cohort study); alternatively, they could calculate the sample size required to achieve a chosen target mean uncertainty interval width, MAPE or misclassification probability.

When performing the calculations, the actual clinical setting of interest should be considered, with input from key stakeholders including patients and health professionals about acceptable levels of individual-level uncertainty and, if relevant, misclassification probabilities. Currently, such discussions between the model developers and stakeholders are either not done at all or tend to be conducted after model development. We hope our approach will motivate communication around this topic to begin at the onset of the development study, in the context of the clinical decision to be made and any relevant risk thresholds. Eliciting the key time-point(s) of interest for prediction is crucial, via discussion with patients, clinicians and other stakeholders.[38] In the breast cancer example, we focused on 5-year risk of recurrence, but sometimes multiple time-points may be of equal interest (e.g., 1-year, 5-year and 10-years), and the sample size required will change depending on the chosen time-point.

Our proposal utilises maximum likelihood estimation theory for unpenalised exponential regression models, and uses Fisher's information matrix to examine epistemic uncertainty (*reducible* model-based uncertainty) that arises from fitting an assumed 'core model' with a core set of predictors.[39] We do not consider aleatoric uncertainty (*irreducible* uncertainty) that refers to residual uncertainty that cannot be explained by the 'core model'. The proposal also assumes parameter estimates will be unbiased (agree with the true 'core model' predictor effects on average). This is why we recommend starting at the minimum recommended by *pmsampsize*, which at least aims to minimise overfitting and aim for well-calibrated predictions at the population level. Also, unpenalised and penalised (shrinkage) approaches are likely to quite closely agree in this situation. Further research is needed to consider more complex settings, for example to deal with large numbers of candidate predictors (e.g., core predictors plus potential others, including noise variables), allowing variable selection, and penalised regression approaches like lasso. Other machine learning approaches, such as tree-based methods, may need substantially higher sample sizes to achieve the same level of stability compared to (penalised) regression approaches.[40] Again, further research is needed to examine this.

We assumed an exponential regression for simplicity, as it makes the 'core model' easier to pre-specify with an assumed constant baseline hazard, and facilitates closed-form solutions that are helpful for a sample size calculation. In two examples (Section 6) the exponential model was a good approximation for quantifying uncertainty of risk estimates relative to either a Weibull model or flexible parametric model, even though the baseline hazard was not constant. Cox states that he generally prefers specifying survival models parametrically as "various people have shown that the answers are very insensitive to the parametric formulation of the underlying distribution".[41]

The main obstacles are specifying the 'core model' and the joint predictor and censoring distributions. Choices may involve some subjectivity, and so sensitivity analyses may be sensible (as is often the case with any sample size calculation). We provided pragmatic suggestions to facilitate the process, for example by focusing on a small number of core predictors (and protected characteristics) of interest, and by basing the 'core model' on previously published models or by assuming equal weighting of (standardised) predictors whilst adhering to a particular overall risk and C-index. Generally, our approach is more easily implemented when a synthetic or existing dataset is available (which is often the situation in practice), as then the joint distribution of core predictors can be observed directly, potentially alongside the censoring and event-time distribution.

Lastly, there is the potential for automating some or all of the sample size calculation steps, by integrated them into existing software platforms that are designed to extract study datasets and process healthcare data, for instance the Data extraction for epidemiological research (DExtER) tool.[42] Alternatively, they could be embedded alongside the production of synthetic datasets or at the point of obtaining feasibility information from large, anonymised longitudinal electronic healthcare record (EHR) databases (e.g., CPRD). This would facilitate and encourage appropriate sample size calculations by those using EHR data to develop prediction models.

In summary, we have proposed a new approach to inform the (minimum) sample size required for developing a clinical prediction model with a time-to-event outcome based on a 'core model' of established predictors. The approach enables researchers to examine how the sample size (for new data collection or an existing dataset) impacts individual-level uncertainty intervals and classification instability, to guide decisions on suitable datasets and sample size targets for model development.

# SUPPLEMENTARY MATERIAL

## S1: How to measure the impact of uncertainty on clinical utility

Alongside the summary statistics and plots in the main paper, we can also measure the impact on clinical utility. If true risks (from the 'core model') ae known, then we have perfect information, and thus remove any prediction uncertainty and potential for misclassification, and maximise clinical utility from the core model. But with a particular sample size, the estimation error could reduce clinical utility, and we measure this as follow:

- *Expected loss in net benefit (plot and summary statistics).* Net benefit is a measure of clinical utility at a risk threshold ($z$) chosen for making decisions (e.g., initiating biopsy, starting treatment); for brevity we refer to explanations elsewhere.[31] Each sampled value (e.g., 1000) of an individual's estimated risk ($q_i$) from their prediction uncertainty distribution, leads to a sampled value of their loss (i.e. absolute difference) in net benefit ($d_i$) compared to knowing their true risk ($p_i$, from the core model) , where:

$$d_i = \begin{cases} 0 & \text{if sign}(q_i - z) = \text{sign}(p_i - z) \\ \left| p_i - \left((1-p_i)\left(\frac{z}{1-z}\right)\right) \right| & \text{otherwise} \end{cases} \quad \text{Eq.(12)}$$

Essentially, $d_i$ is zero when the sampled estimated risk ($q_i$) and the true risk ($p_i$) are either both above or both below the threshold; otherwise $d_i$ is the absolute value of an individual's contribution to the net benefit function. Finally, we can calculate,

$$\text{Expected loss in net benefit for individual } i = \Delta_i = \text{E}(d_i) \quad \text{Eq.(13)}$$

which is estimated by their average sampled value of $d_i$, or equivalently

$$\Delta_i = \text{E}(d_i) = P(\text{misclassification})_i \times \left| p_i - \left((1-p_i)\left(\frac{z}{1-z}\right)\right) \right|$$

This can be plotted on a 'Expected net benefit loss plot', with $\Delta_i$ on the y-axis and $p_i$ on the x-axis. We can also sum $\Delta_i$ across all individuals, to estimate the expected total loss in net benefit in the population ($\Delta$).

For the breast cancer example, the results obtained were as follows:

| Data used to inform sample size calculations; assumed core model | Sample size for new model development | Expected net benefit loss at 20% threshold *mean (min, med, max, sum)* |
|---|---|---|
| GBSG dataset of 220 patients; original core model - see Section 4.1 | 355 (minimum no. recommended by *pmsampsize*) | 0.00003 (0, 0.00003, 0.015, 0.524) |
| | 920 (no. to target 95% uncertainty interval widths ≤ 0.2 in those with true risks ≤ 0.3) | 0 (0, 0, 0.0093, 0.231) |

**S2: Example code for extended analyses**

**(i) to generate a synthetic dataset from a real dataset in R**

```
# load the real dataset, load the synthpop module and choose a seed
library(synthpop)
myseed <- 66

# apply the synthpop module choosing the variables and their suitable assumed
distributions
# here, the dataset only contains all the 7 variables of interest
# visit.sequence arranges the order of the variables
# first variable is always just a random sample of the original dataset so worth examining
impact of this choice
# use minimumlevels to tell it which are factor variables
# k defines the size of the synthetic dataset
mysyn <- syn(myrealdata, minnumlevels = 3, method = c("norm", "lognorm", "polyreg",
"lognorm", "lognorm", "logreg", "sample" ), k = 10000, visit.sequence = c(7, 1, 2, 3, 4, 5, 6),
seed = myseed)
```

### (ii) using survsim in Stata to generate survival times

```
*lambdas set to 0.0000975 based on trial and error to get S(5) to be about 0.39
set seed 560
survsim rectime , distribution(exponential) lambdas(0.0000925) covariates(LP_test 0.208)
gen censrec = 1
stset rectime, failure(censrec==1) scale(365.25)
sts graph

* with the original breast cancer data and swithing the event indicator to be 1 when they are censored
* we find that 10% censoring prob by 2 years, and then 70% by 5 years
* assume no censoring by two years but then uniform censoring afterwards with max follow-up of 7.2 years which is the assumed length of the new study like original GBSG daatset
gen censtime = (2*365.25) + runiform()*(5.28*365.25)
replace censrec = 0 if censtime < rectime
replace rectime = censtime if censtime < rectime
stset rectime, failure(censrec==1) scale(365.25)

* visualise the final KM curve
sts graph

* can check the censoring rate
*replace _d = (-1+_d)*-1
* sts graph
summ _t
streg age_sd size_sd nodes_sd g2 g3 meno , dist(exp) nohr
streg LP_test, dist(exp) nohr
```

**S3: Impact of changing risk threshold and predictor weights in the breast cancer example**

The required sample size depends heavily on the choice of risk threshold, which emphasises the need to establish relevant risk thresholds with stakeholders ultimately involved in the decision-making process. To illustrate this, consider that the designated risk threshold been much lower than 20%, say 5%; then the mean probability of misclassification is almost zero, even with a sample size of 355, as nearly all participants have 'true' risks much greater than 0.05. In that situation, larger sample sizes than suggested by *pmsampsize* may not be as necessary. Conversely, had the threshold been much higher, say 50%, then the mean probability of misclassification is 0.11 with 355 participants, about double that when using 20% threshold.

Let us now consider the impact of changing the 'core model' specification, for example to mimic a situation where less information is available about the relative predictor weights, and so pragmatic decisions are required. Recall in Section 3 we suggested, in the absence of more detailed information, to assume equal weights of *standardised* continuous predictors and set sensible weights for categorical variables. To examine this for the breast cancer model, we modified the 'core model' from Step (3) to be:

$$t_i \sim \text{exponential}(\eta_i)$$
$$\ln(\eta_i) = \mu_i$$
$$= \alpha + \delta((-1 \times \text{age}) + (1 \times \text{size}) + (1 \times \text{\#nodes})$$
$$+ (1 \times \text{post\_menopause}) + (1.5 \times \text{grade2}) + (3 \times \text{grade3}))$$

This is labelled as the 'alternative core model'. Standardised continuous predictors are each given the same weight, but the direction of effect is chosen to reflect previous knowledge that a younger age, a greater tumour size, and a higher number of nodes is associated with a higher recurrence risk. Furthermore, previous knowledge suggests post-menopausal women and a higher tumour grade are at increased risk, with post-menopause given the same weight as the continuous predictors, and tumour grades 2 and 3 assigned the highest relative weights (as grade is a well-known strong prognostic factor). We still assume prior information suggests an overall risk of 0.39 and C-index of 0.70 in the target population, and trial and error identifies that $\alpha$ and $\delta$ values of -2.911 and 0.230, respectively, ensure the 'alternative core model' achieve this in the GBSG dataset.

Applying our sample size approach, the prediction and classification instability plots are shown below. By changing the 'core model', the distribution of estimated risks changes, and there are fewer individuals with true risks below the 20% risk threshold. However, the magnitude of prediction and classification instability is quite similar to the original 'core model', as evident by very similar summary statistics for their uncertainty interval widths and MAPE values. For example, with a sample size of 355 participants, the mean uncertainty interval width is 0.23 for both core models, and MAPE is 0.046 and 0.047. As there are fewer individuals close to the 20% threshold, the mean misclassification probably is slightly lower than before, but otherwise changing the core model (though still retaining sensible weights) has not had a substantial impact. Clearly, the more discrepant the assumed core models, the more potential for differences; the same applies to assumptions about the joint predictor and censoring distributions.

(a) 355 participants

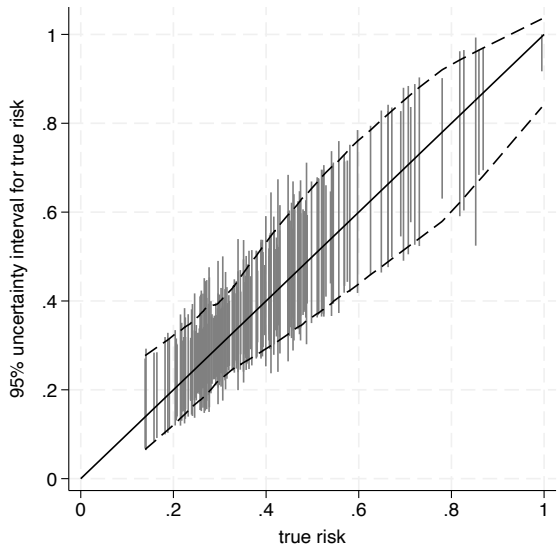
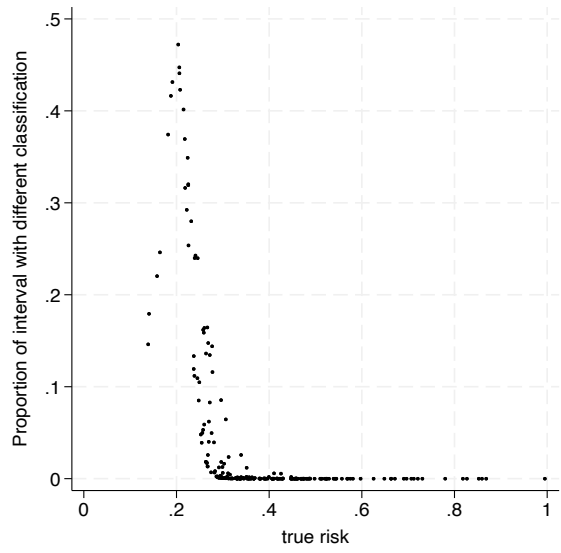

(b) 920 participants

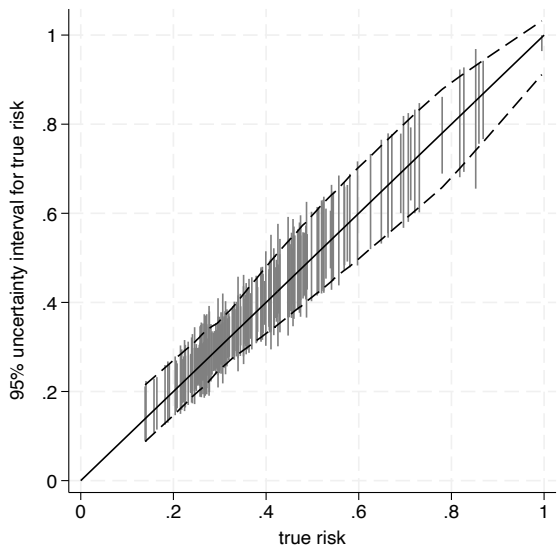
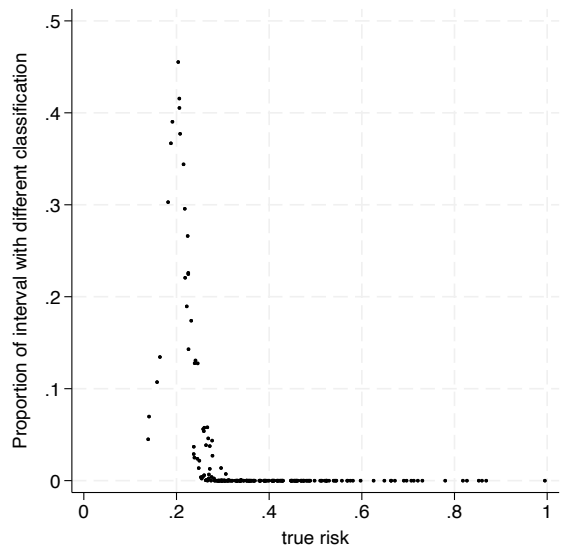

**S4: Comparison of risk estimates and uncertainty interval widths from Weibull and exponential regression models, when applied to the breast cancer dataset of 220 participants**

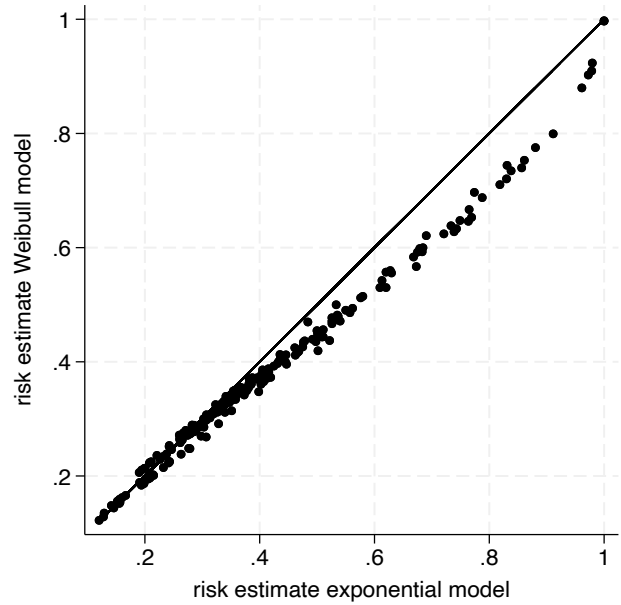

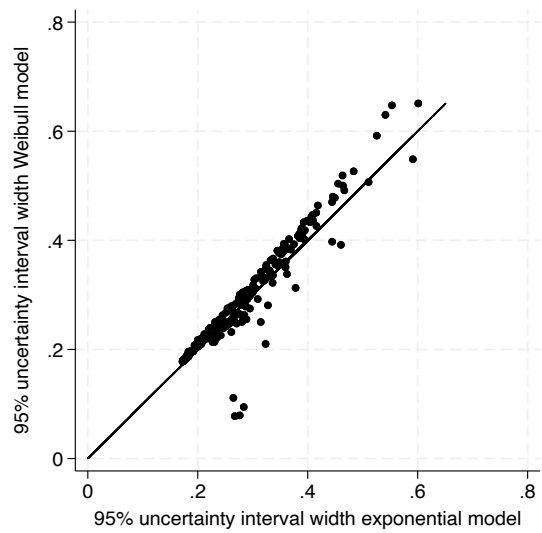

**S5: Comparison of risk estimates and uncertainty interval widths for exponential and Weibull models applied to a prostate cancer dataset involving 502 participants and 4 predictors.**

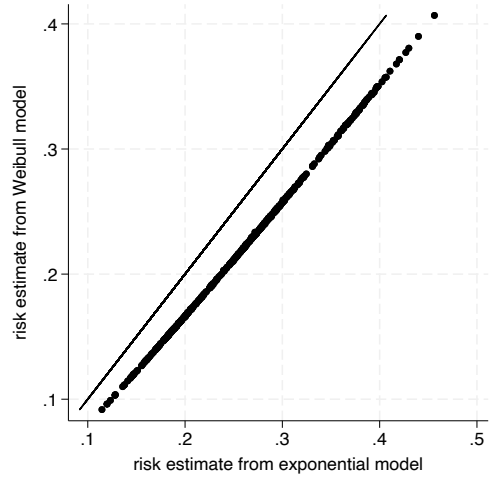

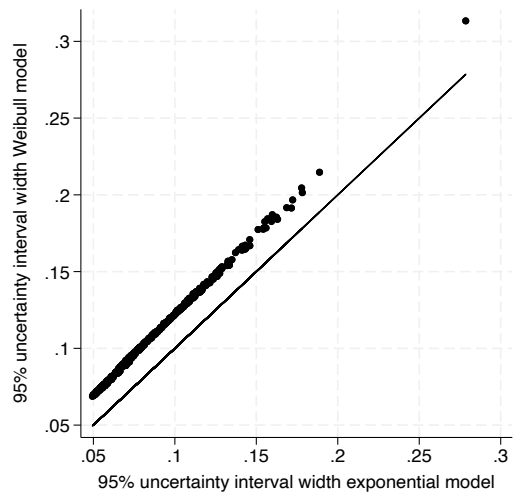